\documentclass{JINST}
\usepackage{lineno}
\title{The Pixel Detector of the ATLAS experiment for the Run2 at the Large Hadron Collider}

\author{Y. Takubo on behalf of the ATLAS collaboration\\
\llap{$^a$} High Energy Accelerator Research Organization (KEK),\\
  1-1 Oho Tsukuba Ibaraki, 305-0801, Japan\\
E-mail: \email{yosuke.takubo@kek.jp}}

\abstract{The Pixel Detector of the ATLAS experiment has shown excellent performance during the whole Run-1 of LHC. Taking advantage of the long shutdown, the detector was extracted from the experiment and brought to surface, to equip it with new service quarter panels, to repair the modules and to ease installation of the Insertable B-Layer (IBL). The IBL is a fourth layer of pixel detectors, and has been installed in May 2014 between the existing Pixel Detector and a new smaller radius beam-pipe at a radius of 3.3 cm. To cope with the high radiation and pixel occupancy due to the proximity to the interaction point, a new read-out chip and two different silicon sensor technologies (planar and 3D) have been developed. Furthermore, the physics performance will be improved through the reduction of pixel size while, targeting for a low material budget, a new mechanical support using light weight staves and CO$_{2}$ based cooling system have been adopted. An overview of the refurbishing of the Pixel Detector and the IBL project as well as the experience in its construction will be presented, focusing on adopted technologies, module and stave production, qualification of assembly procedure, integration of staves around the beam pipe and commissioning of the detector.}

\keywords{ATLAS; Pixel Detector; IBL}

\begin{document}
\section{Introduction}\label{sec:intro}
\begin{figure}[bp] 
\centering
\includegraphics[width=.6\textwidth]{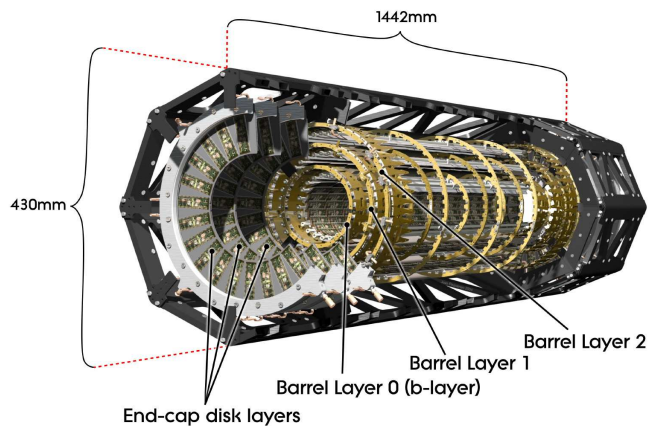}
\caption{Schematic drawing of the ATLAS Pixel Detector. The detector consists of three barrel layers and two endcaps with three disc layers each.}
\label{fig:pixel_detector}
\end{figure}

The ATLAS Pixel Detector \cite{pixel} is the innermost tracking detector of the ATLAS experiment \cite{atlas}. It consists of three barrel layers placed at the radius of 50.5 mm, 88.5 mm and 122.5 mm centered around the beam axis ($z$) and 2 endcaps with three disc layers each (Fig.~\ref{fig:pixel_detector}). The full detector contains 1744 modules, and the total number of the pixels is 80 million.

Each Pixel module uses a n$^{+}$-in-n silicon planar sensor with 47232 pixels whose typical size is $50 \times 400$ $\mu$m$^2$ and thickness is 250 $\mu$m. A sensor of the module is bump-bonded with 16 front-end chips (FE-I3 chips). A Module Control Chip (MCC) is mounted on each Pixel module to control 16 FE-I3 chips on it.

During the two years of LHC run between 2011 and 2012 (Run-1), the Pixel Detector showed stable and good performance. The efficiency for a track passing through the Pixel Detector was about 99\% for non-disabled modules, and the data taking efficiency was 99.9\%. The typical noise value of the Pixel modules is about 160 electrons with the tuning of the threshold value of 3500 electrons.

LHC was stopped in February 2013 for a shutdown of two years (LS1: Long Shutdown 1) to upgrade the collision energy to 13 TeV with an instantaneous luminosity of about $1 \times 10^{34}$ cm$^{-2}$s$^{-1}$. During LS1, two upgrade programs were performed for the Pixel Detector. One is the replacement of the Service Quarter Panel (SQP) and the other is the installation of the Insertable B-Layer (IBL). In this paper, the overview of these upgrade programs is described.

\section{Replacement of service panel of Pixel Detector}\label{sec:pixel}

In the Pixel Detector, the opto-boards are used for the electric-optic conversion of the signals from the modules. In Run-1, the opto-boards were housed on the SQPs which were installed with the detector and carry electrical power, cooling and optical data both into and out of the detector. Since the opto-boards were put inside the detector region, they were not accessible for repairs. In order to make them more accessible for the potential repairs, the Pixel Detector was extracted from the ATLAS detector in April 2013 to replace SQPs by New Service Quarter Panels (nSQP) during LS1. 

The nSQP extends the electrical lines with 6.6 meter to the outside of the detector after the electrical bundles of 1 m long which are connected with the Pixel modules. Since the signals from the modules are not strong enough to be transmitted for such length, electrical repeater chips are mounted on nSQP to boost the signals. With the nSQP, the electric lines can be extended to the outside of the detector, and it realizes the easier access to the opto-boards. Even if a problem happens at the opto-boards, it can be fixed easily without extracting the detector.

Another problem occurring in the Pixel Detector during Run-1 was a high number of desynchronized modules due to limitation of the bandwidth in the optical fiber between opto-board and the readout electronics. Although the data transmission speed from a MCC is 160 Mbps for the innermost layer (B-Layer) by using two optical lines whose bandwidth is 80 Mbps, the maximum speed was 80 Mbps for the second layer (Layer-1) because only one optical line was installed for them. Extra optical lines were installed for Layer-1 during LS1 to realize the data transmission speed of 160 Mbps.

\begin{figure}[tbp] 
\centering
\includegraphics[width=.6\textwidth]{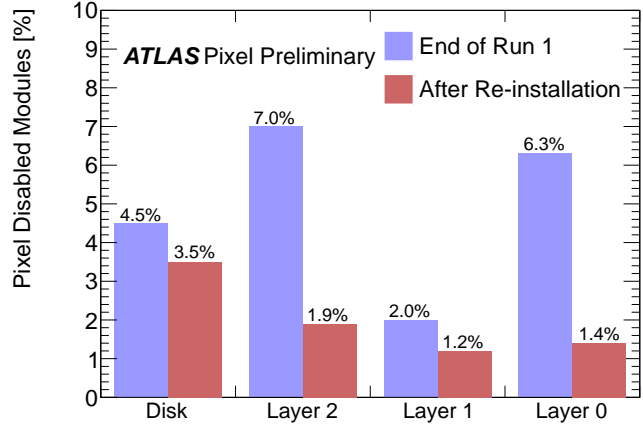}
\caption{The number of the disabled Pixel modules at the end of Run-1 and after re-installation of the Pixel Detector with the nSQPs into the ATLAS detector \cite{pixel_official}. Layer-0 in the figure means B-Layer.}
\label{fig:pixel_disabled_module}
\end{figure}

At the beginning of Run-1, the fraction of disabled modules in the Pixel Detector was 2.5\%, and then it increased gradually during operation. The fraction of the disabled modules became 5\% at the end of the Run-1. Once the Pixel Detector was moved to the ground, the non-operational modules were investigated. There were several sources of problems like broken front-end chips and disconnection of the electrical lines in SQP. They were fixed as much as possible during replacement of the SQPs. The Pixel Detector was re-installed into the ATLAS detector in December 2013. Figure~\ref{fig:pixel_disabled_module} shows the fraction of the disabled modules in each layer of the Pixel Detector at the end of the Run-1 and after the re-installation. The disabled modules were reduced to 1.9\% after the re-installation \cite{pixel_official}.

\section{Installation of the IBL}\label{sec:ibl}
\subsection{Overview}

The instantaneous luminosity of LHC will be upgraded to about $2 \times 10^{34}$ cm$^{-2}$s$^{-1}$ after LS2 which is planned in 2018. The large amount of the hit data from the Pixel Detector will become a serious issue for the front-end electronics in the operation after LS2 since it was designed to work at the instantaneous luminosity up to $1 \times 10^{34}$ cm$^{-2}$s$^{-1}$. Especially, the data transmission speed and buffer size of the MCC are not enough to cope with the data obtained at the instantaneous luminosity of $2 \times 10^{34}$ cm$^{-2}$s$^{-1}$. If, however, a new pixel layer is installed inside the inner most layer of the Pixel detector, B-Layer, it can recover the efficiency drop of the Pixel Detector and improve the performance. For that reason, a new pixel layer (IBL) \cite{ibl} was installed inside B-Layer during LS1. In the following, when comparing the technologies of the IBL with the pixel detector used for Run-1, the latter will be referred as the Pixel Detector.

\subsection{IBL module production} \label{sec:module}
The modules of the IBL adopt two different sensor technologies: n$^{+}$-in-n planar sensors and 3D sensors. The main characteristics of these sensors are summarized in Table~\ref{tb:sensor}. The pixel size of the sensors is $50 \times 250$ $\mu$m$^{2}$ which is 60\% of the pixel size used for the Pixel Detector. The IBL has about 12 million pixels in total. The thickness of the planar and 3D sensors is 200 $\mu$m and 230 $\mu$m, respectively. The sensors were designed to work at a radiation dose up to $5 \times 10^{15}$ n$_{\mathrm{eq}}$ which is the expected dose by the end of LHC Phase-1 operation around 2022.

\begin{table}[tbp]
\caption{The main characteristics of n$^{+}$-in-n planar sensor and 3D sensor used for the IBL.}
\label{tb:sensor}
\smallskip
\centering
\begin{tabular}{|l|r|r|} \hline
& Planar sensor & 3D sensor \\ \hline
Pixel size ($\mu$m$^{2}$) & $50 \times 250$ & $50 \times 250$ \\ \hline
Sensor size (mm$^{2}$) & $41.32 \times 18.59$ & $20.45 \times 18.75$ \\ \hline
Thickness ($\mu$m) & 200 & 230 \\ \hline
Bulk type & n-type & p-type \\ \hline
\end{tabular}
\end{table}

n$^{+}$-in-n planar sensor is a well developed technology and it is the one used for the Pixel Detector. n$^{+}$-in-n planar sensor uses n-type bulk, and  n$^{+}$ implants and p$^{+}$ implants are put as the electrodes at the sensor surface and the side of the bump-bonding with a front-end chip to apply the high voltage (HV) and ground, respectively. 

The edge around the guard-ring is inactive in planar sensor. The sensors used for the Pixel Detector have an inactive region of 1100 $\mu$m from the edge, therefore, Pixel modules are placed with overlaps along the beam axis to compensate this inactive region. Since the space of the radial axis ($r$) is limited in the case of the IBL, it is impossible to overlap the modules. For that reason, planar sensors were developed with a smaller inactive region of 200 $\mu$m, extending the pixels at the sensor edge under the guard-ring \cite{ibl_module}.

In 3D sensors, n$^{+}$ and p$^{+}$ implants are embedded vertically into the p-type bulk as the electrodes to apply the HV and ground, respectively. Since the electrodes are put closely with a distance of 50 $\mu$m, the 3D sensor can be fully depleted with much lower HV, compared to a planar sensor. For 3D sensors used for the IBL, 6 V is enough for the full depletion before the irradiation. It was designed to have 200 $\mu$m inactive region at the sensor edge. The IBL is the first large scale application of this new technology.

A new front-end chip (FE-I4B chip) was developed for the IBL. The FE-I4B chip was produced with IBM 130 nm technology, and $80 \times 336$ pixels are embedded per chip. It is designed to have a radiation hardness up to 250 Mrad. In FE-I4B chip, hit data in each pixel are stored locally, and the pixel matrix is organized in four pixel digital readout units which provide local hit processing and distribute a buffer to store hit data until receiving a trigger. The signals from the sensor are amplified and digitized in the analog and digital circuits respectively, and finally the charge information is output with 4-bit value of TOT (Time-Over-Threshold).

The cost of the bump-bonding is not proportional to the chip size but to the number of chips. For that reason, the front-end chip with large size reduces the cost of the bump-bonding. The FE-I4B chip was, therefore, designed to have a large area of $20.2 \times 18.8$ mm$^{2}$. The size is 4.5 times larger than that of the FE-I3 chip which is used for the Pixel Detector.

The sensors and FE-I4B chips are connected with bump-bonding by using AgSn, and then the flexible PCB is glued on the sensor side to make a module. Two FE-I4B chips are attached on a planar sensor and one chip is used for a 3D sensor, therefore, the size of a planar module is two times larger than that of a 3D module. 

\begin{figure}[tbp] 
\centering
\includegraphics[width=.8\textwidth]{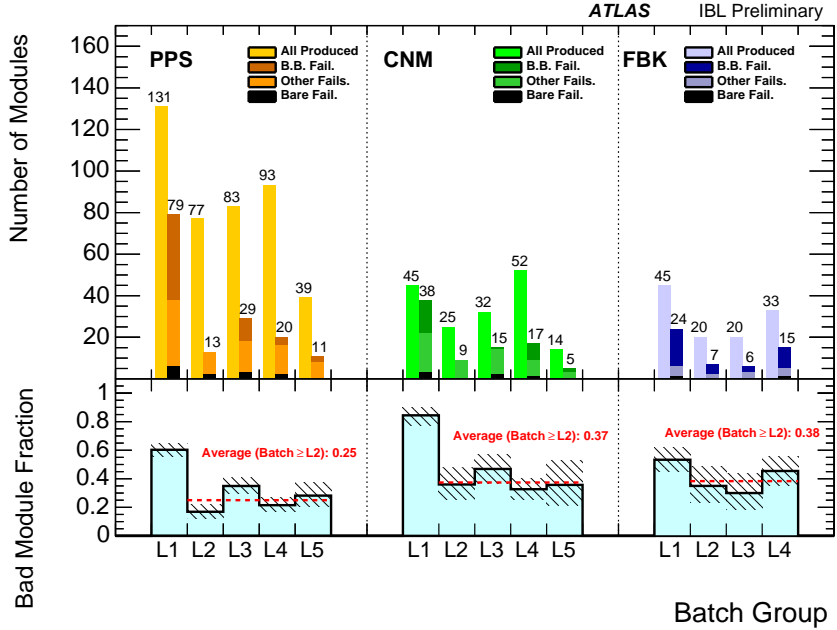}
\caption{The number of the produced modules (top) and the fraction of the bad modules (bottom) per production batch group for planar (PPS) and 3D sensors (CNM and FBK) \cite{pixel_official}. FBK and CNM are two vendors of 3D sensors.}
\label{fig:module_yeild}
\end{figure}

Figure~\ref{fig:module_yeild} shows the number of the produced modules and the fraction of the bad modules \cite{pixel_official}. At the beginning of the production (L1 in Fig.~\ref{fig:module_yeild}), open and merged bumps were found in a high rate of the modules. One of the reasons of this issue was to use too much flux during the flip-chip process. The problem was solved after adopting flux-free flip-chip process. The total production yield was 75\% for the planar module and 62\% for the 3D module after solving the bump issues.


\subsection{IBL stave production}
\begin{figure}[tbp] 
\centering
\includegraphics[width=.45\textwidth]{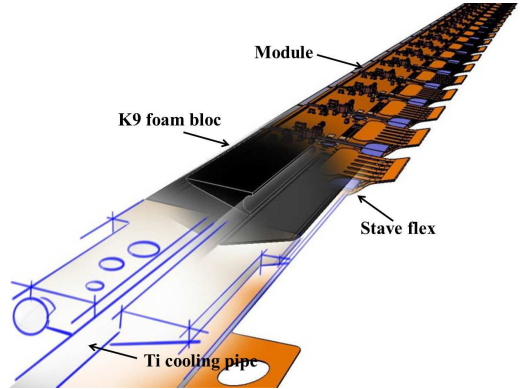}
\caption{Schematic view of the IBL stave.}
\label{fig:stave}
\end{figure}

Figure~\ref{fig:stave} shows a schematic view of the IBL stave. A bare IBL stave is made of carbon foam with the size of $74.8 \times 1.88$ cm$^2$ and its cross-section has a triangular shape. Inside a IBL bare stave, a cooling pipe with 1.5 mm diameter is inserted, which is made of titanium. The planarity of the $r$-direction is required to be within 350 $\mu$m to put the staves in the limited space between the beam-pipe and the B-Layer of the Pixel Detector. The material budget of the stave is only 0.6\% ($X/X_{0}$) including the cooling pipe. 

Two flexible PCBs (stave flex) are glued on the two sides of $z$-direction of the bare stave to provide the electrical lines. In the stave flex, the low voltage (LV) lines for the FE-I4B chips are made of aluminum to reduce the material budget.

On each stave, 12 planar modules are mounted in the middle region in $z$-direction covering 75\% area and 8 3D modules in the edge region covering 25\% area. As described in Sec. \ref{sec:module}, there is not enough space to put the modules with overlap in  $z$-direction for the IBL, so that the inactive region of $z$-direction must be reduced as much as possible. For that reason, the modules were mounted with a gap less than 350 $\mu$m along $z$-direction. The stave flex and modules were glued together, and then all the electrical lines were connected with wire bonding.

During the production of the staves, Al(OH)$_{3}$ was found on the bonding pad on 13 staves, which comes from corrosion of bonding wires made of aluminum. Al(OH)$_{3}$ can be produced with DI water, and besides, the corrosion process is accelerated with halogen elements like chlorine and fluorine. The water came from condensation during thermal cycling where temperature was changed from $-40$ to $40$ degrees to investigate thermal tolerance of the staves. Chlorine and fluorine were also found on the bonding pad, and they seem to accelerate the corrosion process. The source of chlorine and fluorine, however, was not identified eventually. To avoid the corrosion, the humidity control was improved in the staves testing. 

\begin{figure}[tbp] 
\centering
\includegraphics[width=.6\textwidth]{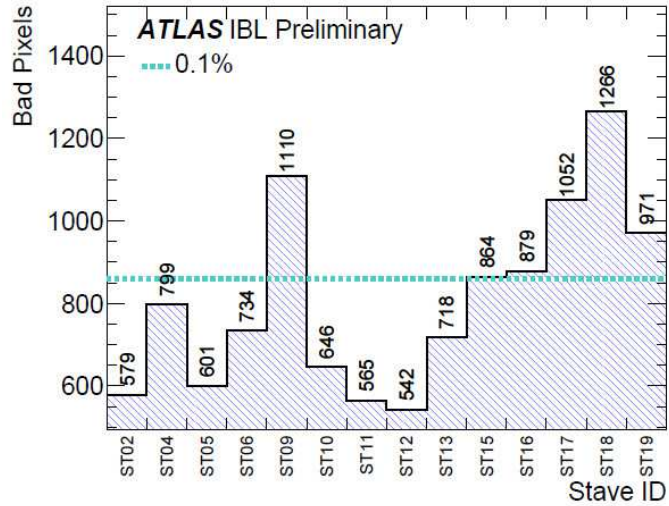}
\caption{The number of the bad pixels in the 14 staves used for IBL \cite{staveqa}.}
\label{fig:stave_badpixel}
\end{figure}

Once the staves were produced, electrical functionality tests were performed to investigate their performance in detail. Any possible damage of the wire bonding and module surface was checked by optical inspection with a microscope. The quality of the sensors was confirmed with Current-Voltage (I-V) measurement. A breakdown voltage above 80 V and 20 V was required for the planar sensor and the 3D sensor, respectively. The functionality of the FE-I4B chips was checked by configuring the chips and reading data from them, including the tuning of the chip parameters. Finally, source scans with $^{241}$Am and $^{90}$Sr were performed to test the overall functionality of modules. In these tests, it was confirmed that IBL can be operated with a threshold value of 1500 electrons with noise values below 200 electrons. Figure~\ref{fig:stave_badpixel} shows the number of the bad pixels in the 14 staves used for the IBL \cite{staveqa}. The fraction of bad pixels is required to be below 1\% in this test, and all the staves well satisfied this requirement. 

\subsection{Integration and installation of IBL}
The integration of IBL started on February, 2014. The beam pipe used in Run-1 was replaced by a new one with an inner radius reduced from 29 mm to 23.5 mm. The IBL was integrated around the new beam pipe and inserted into the ATLAS detector together with it.

Before the integration, the cooling pipes in the staves were extended to 7 m in length by brazing. The cooling is one of the critical parts in the IBL: the sensors must be cooled down to $-40$ degrees due to the high radiation damage (The radiation dose on the sensors is expected to be $5 \times 10^{15}$ n$_{\mathrm{eq}}$ at the maximum). Although C$_{3}$F$_{8}$ has been used for the Pixel Detector whose cooling capability is up to $-15$ degrees, its cooling power is not enough for the IBL. For that reason, CO$_2$ evaporative cooling is adapted for the IBL, since it has higher cooling efficiency which allows to reduce the diameter of the cooling pipe and then the material.
	
After brazing the cooling pipes, the 14 staves were put on the support structure on the IBL Support Tube (IST) which is located around the beam pipe to integrate the staves. Then, the service cables, about 5 m long, were connected at each edge of the staves to provide the data lines, HV and LV. Two or three staves were integrated on the IST every week, and the integration finished in two months in March 2014.

After the integration, the test system was connected with the service cables, and an electrical functionality test was performed through the service cables to find any damage of the detector during the integration. In this test, one shorted line and disconnected line were found in the service cables, and they were fixed by replacing the cables. Except for that, no damage and degradation of the performance were found.

\begin{figure}[tbp] 
\centering
\includegraphics[width=.45\textwidth]{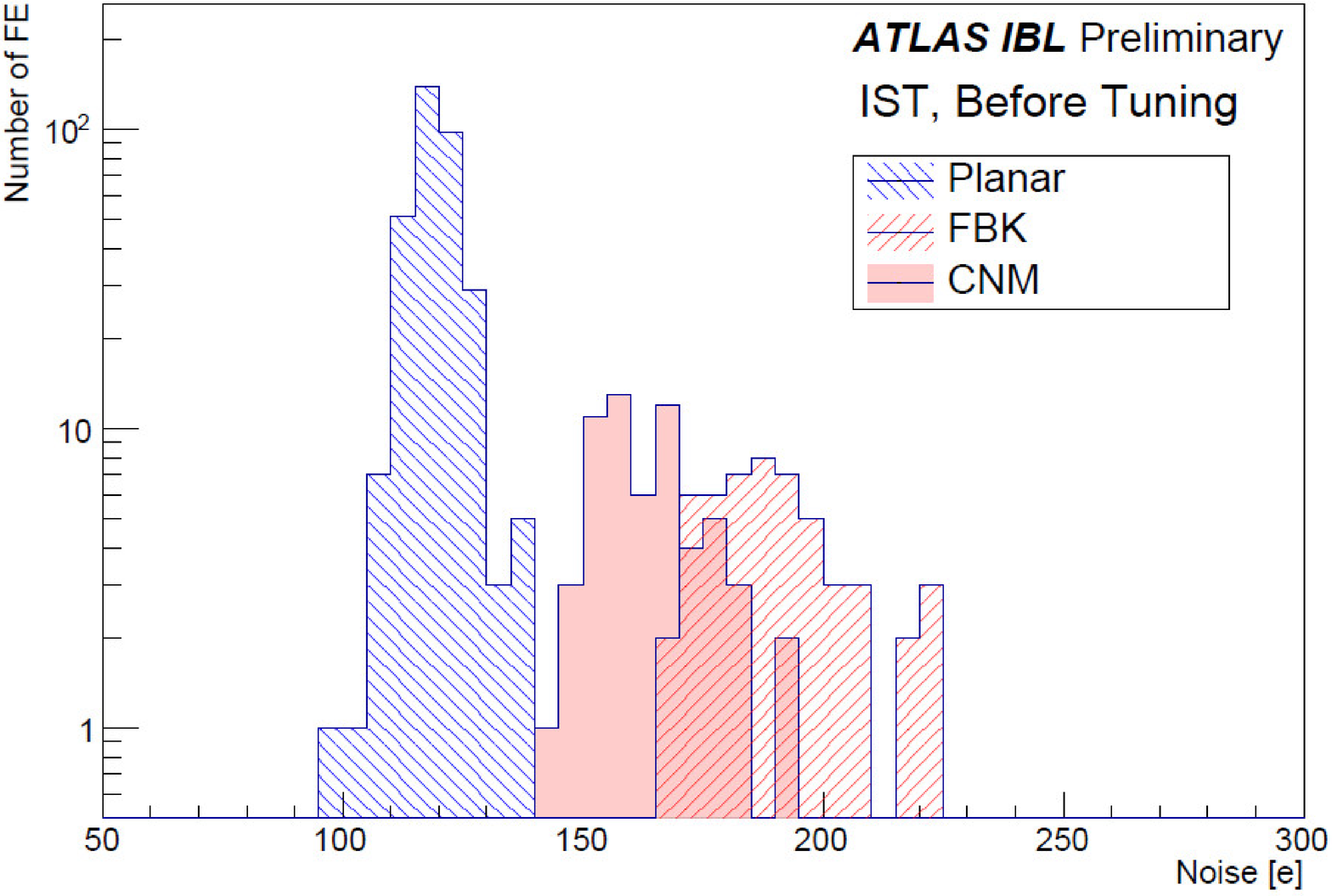}
\includegraphics[width=.45\textwidth]{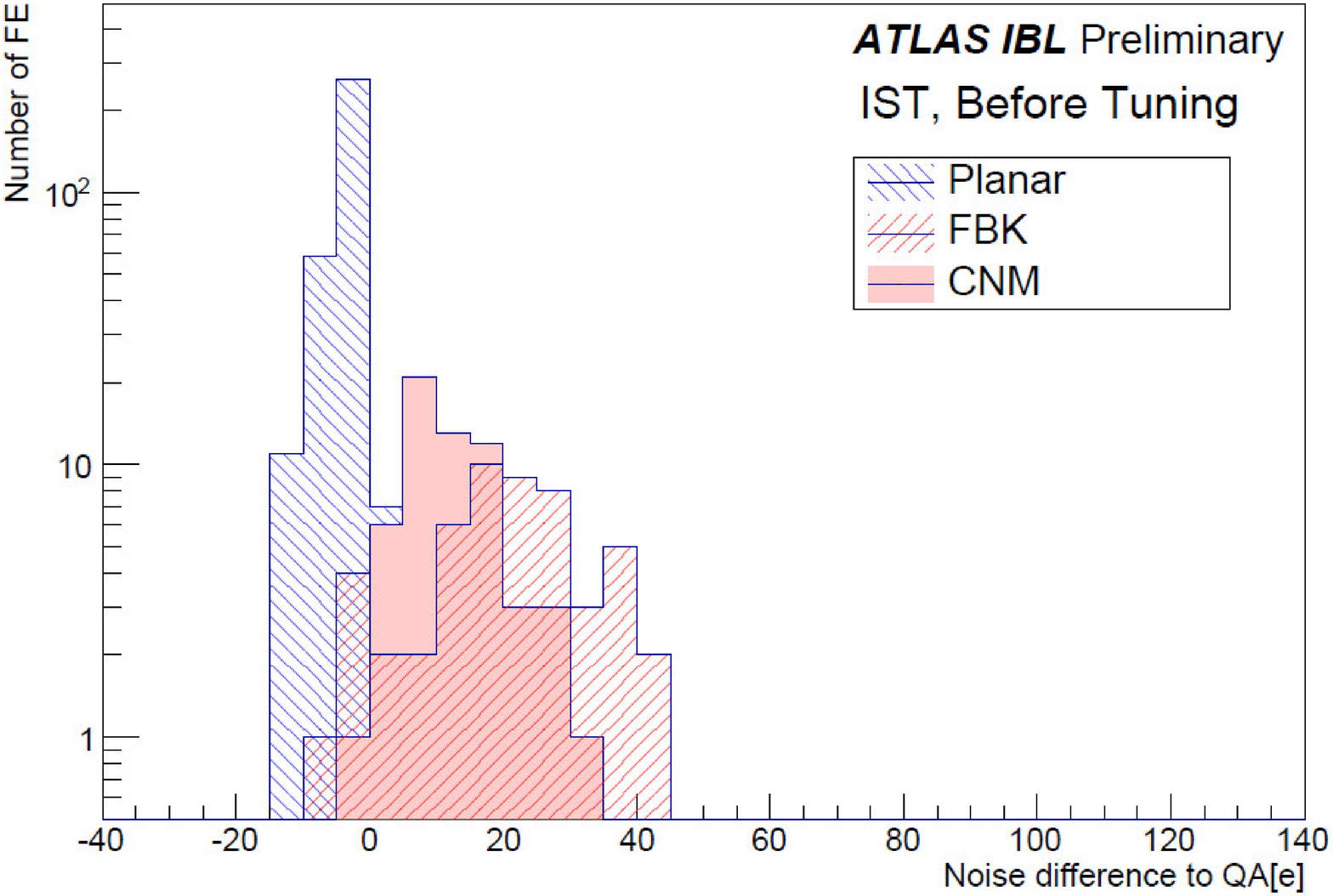}
\caption{The noise distribution obtained after installation of the IBL (left) and the difference of the noise values from the results obtained before integration of the staves on the IST (right) \cite{econ}.}
\label{fig:econ_noise}
\end{figure}

The IBL was transported to the ATLAS experimental hall at 90 m underground on May 5, 2014 and installed into the ATLAS detector on May 7. After connecting the cooling pipes and service cables for the powering and environmental monitors, the performance of the IBL was tested again with the test system. In this test, the final DCS system was used for powering and reading the environmental monitors. The test system was connected with the service cables of the data lines to read-out the modules on the staves. Figure~\ref{fig:econ_noise} shows the difference of the noise values between this test and the stave quality assurance test performed before integration of the staves \cite{econ}. In this test, it was confirmed that 100\% of the sensors and FE-I4B chips is operational. Besides, there was no significant change of the performance from the stave quality assurance test.

\section{Summary}\label{sec:summary}
The Pixel Detector of the ATLAS experiment has shown excellent performance during the whole Run-1 of LHC. During LS1, the detector was extracted from the experiment and brought to the surface to equip it with nSQPs and repair the disabled modules. The disabled modules were reduced from 5\% to 1.9\% after the re-installation of the Pixel Detector.

In addition to the replacement of SQPs by nSQPs, the IBL was installed into the ATLAS detector in May 2014 as the new fourth layer of the Pixel Detector. The IBL can be operated with a threshold value of 1500 electrons with noise values below 200 electrons. All the staves well satisfied the requirement of having a fraction of bad pixels less than 1\%. In the test after installation of the IBL, it was confirmed that 100\% of the sensors and FE-I4B chips are operational.

The commissioning of the Pixel Detector and IBL is ongoing to make them ready for Run-2 which will start in spring 2015.







\begin{thebibliography}{9}

\bibitem{pixel}
G.~Aad, et al,
\emph{ATLAS pixel detector electronics and sensors},
\jinst{3}{2008}{P07007}.

\bibitem{atlas}
ATLAS Collaboration,
\emph{The ATLAS experiment at the CERN Large Hadron Collider},
\jinst{3}{2008}{S0800}.

\bibitem{pixel_official}
ATLAS Collaboration, Approved Pixel Plots, http://twiki.cern.ch/twiki/bin/view/AtlasPublic/ApprovedPlotsPixel.

\bibitem{ibl}
ATLAS Collaboration,
\emph{ATLAS Insertable B-Layer Technical Design Report},
CERN-HLCC-2010-013, ATLAS TDR \textbf{19}, 15 September 2010.

\bibitem{ibl_module}
ATLAS IBL Collaboration,
\emph{Prototype ATLAS IBL modules using the FE-I4A front-end readout chip},
\jinst{7}{2012}{P11010}.

\bibitem{staveqa}
ATLAS Collaboration, IBL Stave Quality Assurance, ATL-INDET-PUB-2014-002, 2014, http://cds.cern.ch/record/1707948.

\bibitem{econ}
ATLAS Collaboration, IBL Stave Quality Assurance, ATL-INDET-PUB-2014-036, 2014, http://cds.cern.ch/record/1750215.




\end{thebibliography}
\end{document}